\documentstyle[12pt,twoside]{article}
\begin{document}
\begin{center}

{\Large\bf
A  Quantum Cosmological Model With\\[5PT]
Static and Dynamic Wormholes\\[5PT]}
\medskip

{\bf  N.A.
Lemos\footnote{e-mail: nivaldo@if.uff.br} and G. A.
Monerat\footnote{e-mail:
monerat@if.uff.br}}  \medskip

Departamento de F\'{\i}sica, Universidade Federal Fluminense\\
CEP 24210-340, Niter\'oi, Rio de Janeiro, Brazil
\medskip

\end{center}

\begin{abstract}
Quantization is performed of a  Friedmann-Robertson-Walker  universe  filled with
 a  conformally invariant scalar field and a perfect fluid with equation of state $p=\alpha \rho$.
A well-known discrete set of static quantum wormholes is shown to exist  for radiation ($\alpha =1/3$),
 and
a novel continuous set is found for
cosmic strings ($\alpha = -1/3$), the latter states having throat radii
of any size. In both cases
wave-packet solutions to  the  Wheeler-DeWitt equation are obtained with all the properties
of evolving quantum wormholes.
In the case of a radiation fluid, a  detailed analysis of the  quantum dynamics is made
in the context of the Bohm-de Broglie interpretation.  It is shown that a repulsive
quantum force inversely proportional to the cube of the scale factor
prevents singularities in the quantum domain. For the states considered,
there are no
 particle horizons either.
\vspace{0.5cm}

PACS number(s): 04.20.Cv., 04.20.Me
\vspace{0.5cm}


\end{abstract}

\section{Introduction}

Quantum cosmology and  speculations about quantum effects in the very early Universe began with the work of DeWitt \cite{DeWitt}. Since then,
many investigators have been facing the difficult task of giving a reasonable description of the Planck era.
Friedmann-Robertson-Walker (FRW) models have been intensely studied, and it has been found that
some of them possess wormhole quantum states.
Quantum wormholes are  solutions to the Wheeler-DeWitt equation
which remain regular as the three-geometry collapses to zero and are exponentially damped for
large three-geometries \cite{Hawking}.
In quantum cosmology wormhole states  are not restricted to
exotic types of matter fields, and in fact they exist in models with a simple matter content, such as dust and a
conformally coupled scalar field \cite{flavio2}.

Recently, in collaboration with J. C. Fabris and F. G. Alvarenga, we quantized a FRW model having as matter content a perfect fluid with an arbitrary barotropic equation of
state  \cite{alvarenga}. Wave
packets were constructed
and the behaviour of the scale factor discussed according to the many-worlds and the Bohm-de Broglie
interpretations. Since fundamental fields are believed to play an essential role in the
 dynamics of the early Universe, in the present
 paper we
study the consequences of adding a conformal scalar field to the matter content.
 The treatment  relies on Schutz's canonical formalism \cite{Schutz},
 which  describes a relativistic fluid interacting with the
 gravitational field and
endows the fluid with dynamical degrees of freedom.
The quantum  properties of the model are investigated on the basis of the associated Wheeler-DeWitt equation.
The super-Hamiltonian constraint turns out to be  linear in one of the momenta, and a  time variable can be
naturally introduced,
reducing the  Wheeler-DeWitt equation  to a bona
fide Schr\"odinger equation. The presence of the conformal scalar field gives rise to significant
changes in the quantum dynamics. A well-known discrete set of static wormhole wave functions exists
in the case of
radiation. In the case of   cosmic strings
 a new continuous family of quantum wormhole states is found whose throat radii
can be arbitrarily small or arbitrarily large.
Both for cosmic strings and radiation, wave packets are obtained with the properties of
evolving quantum wormholes.
For a  radiation fluid we manage to construct
the exact propagator, which allows us to study the evolution of initial states of the Gaussian type.
We show that a strong repulsive quantum force, which is inversely proportional to the cube of the scale
factor, prevents the appearence of singularities. There is no particle horizon either.

This paper is organized as follows. In Section 2 a Hamiltonian treatment is given to a FRW universe
filled with a perfect fluid (with an
arbitrary barotropic equation of state $ p=\alpha\rho $)
and a conformally invariant scalar field. In Section 3 the classical equations
of motion are solved for some particular
values
of $\alpha$. In Section 4 the Wheeler-DeWitt equation is constructed. Static quantum wormholes
are obtained in Section 5 for radiation and cosmic strings.
In Section 6   dynamic wormholes are obtained for cosmic strings and radiation, in the latter case by means
of the exact
propagator to the Wheeler-DeWitt equation.
The rest of the paper deals only with
radiation.
The Bohm-de Broglie interpretation is employed in Section 7 to
study the quantum dynamics, and
the role of the quantum force to prevent singularities is highlighted.
Section 8 is dedicated to final comments.

\section{FRW Model With Perfect Fluid and Conformal Scalar Field}

\hspace{0.6cm}A homogeneous and isotropic cosmological model
is characterized by the Friedmann-Robertson-Walker metric

\begin{equation}
ds^2=-N(t)^2a(t)^2dt^2 +  a(t)^2\sigma_{ij}dx^{i}dx^{j},
\end{equation}

\noindent where ${\sigma}_{ij}$ denotes the   metric for a 3-space
of constant curvature $k= +1, 0$ or $-1$, and
the lapse function has been conveniently parametrized as $N(t)a(t)$.

In units such that $c=16\pi G=1$, the  pure gravitational action
 is

\begin{equation}
\label{gravitationalaction}
S_{g}=\int_{M}^{}d^4 x \sqrt{-g}\hspace{0.1cm}R  + 2\int_{\partial M}^{} d^3 x \sqrt{h}\hspace{0.1cm}
K\hspace{0.3cm},
\end{equation}

\noindent where $K$ is the trace of the extrinsic curvature $K_{ij}$ of the boundary $\partial M$ of the space-time manifold $M$.
The matter content is  a perfect fluid plus  a scalar field conformally coupled to gravity. The action associated with the sources of gravity  is

\begin{equation}
\label{af}
S_{m}=\int_{M}^{}d^4 x \sqrt{-g}\hspace{0.1cm}p
- \frac{1}{2}\int_{M}^{}d^4 x \sqrt{-g}\hspace{0.1cm}\bigg(\partial_{\mu}\phi\partial^{\mu}\phi +
\frac{1}{6}R\phi^2\bigg) -\frac{1}{12}\int_{\partial M}^{}d^3 x \sqrt{h}\hspace{0.1cm}K\phi^2\hspace{0.3cm}.
\end{equation}

\noindent
Schutz's canonical formalism \cite{Schutz} makes use of a representation  for the four-velocity of the fluid
as $\,\, U_{\nu} =
\frac{1}{\mu}
\hspace{0.15cm}({\epsilon ,}_{\nu} + \zeta {\beta ,}_{\nu} + \theta {S ,}_{\nu})\hspace{0.15cm}$,
where $\epsilon, \zeta, \beta, \theta, S$ are  five velocity potentials.
The   specific enthalpy $\mu$  is expressed in terms of the velocity potentials by means of the
normalization condition $U^{\nu}U_{\nu}=-1$. The potential $S$ is the specific entropy,
and in  FRW models the potentials $\zeta$ and $\beta$ are zero.

Compatibility with
the homogeneous spacetime metric is guaranteed
by taking the scalar field and  the velocity potentials  as functions of $t$ only. Taking
$p = \alpha\, \rho$ as  equation of state for the fluid,
and
performing an ADM reduction described in detail in \cite{Rubakov,lemos4}, we can write the
total action $\,S_T =S_{g} + S_{m}\,$ as \cite{Rubakov,lemos4,Peleg}

\begin{equation}
S_T=\int_{}^{} dt (p_{a}\dot{a} + p_{\Phi}\dot{\Phi} + p_{\epsilon}\dot{\epsilon} +  p_{S}\dot{S} - N\cal{H})\hspace{0.3cm},
\end{equation}

\noindent where
we have used the constraint $p_{\theta} = 0$ and  redefined the scalar field by means of $\Phi=a \phi$.
The super-Hamiltonian $\cal{H}$ is given by

\begin{equation}
\label{superH1}
{\cal H}=-\frac{p_a^2}{24} - 6ka^2 + p_{\epsilon}^{\alpha + 1}a^{1-3\alpha}\hspace{0.12cm}e^S +
\frac{p_{\Phi}^2}{24} + 6k{\Phi}^2\,\,\, .
\end{equation}

The canonical transformation \cite{alvarenga}

\begin{equation}
\label{canonical}
T=p_S e^{-S}p_{\epsilon}^{-(\alpha+1)}\, ,\hspace{0.3cm} p_{T}=e^{S}p_{\epsilon}^{\alpha + 1}\, ,\hspace{0.3cm}
{\bar\epsilon}=\epsilon- (\alpha + 1)\frac{p_S}{p_{\epsilon}}\, , \hspace{0.3cm} p_{\bar\epsilon}=p_{\epsilon}
\end{equation}

\noindent reduces the super-Hamiltonian   to

\begin{equation}
\label{superH2}
{\cal{H}}= - \frac{p_{a}^2}{24} + \frac{p_{\Phi}^2}{24} - 6ka^2 +6k{\Phi}^2 +
a^{1-3\alpha }p_{T}\hspace{0.3cm}.
\end{equation}
Since $\, {\bar\epsilon}\,$ and $\, p_{\bar\epsilon}\,$ do not appear in the
super-Hamiltonian (\ref{superH2}), they do not represent a dynamical degree of freedom
and may be simply dropped from the action, which  takes the reduced form

\begin{equation}
S=\int_{}^{} dt (p_{a}\dot{a} + p_{\Phi}\dot{\Phi} + p_T \dot{T} - N\cal{H})\hspace{0.3cm}.
\end{equation}

\section{The Classical Equations of Motion}

\hspace{0.6cm}The variational principle $\, \delta S=0\,$ leads to the classical equations of motion

\begin{equation}
\begin{array}{ccc}
{\dot a}=-\frac{N}{12}\,p_a \,\,\,\,\, , \,\,\,\,\, {\dot \Phi}= \frac{N}{12}\,p_{\Phi}
 \,\,\,\,\, , \,\,\,\,\, {\dot T}= N\,a^{1-3\alpha }\hspace{0.3cm},\\
& \\
{\dot p_a}=N\,\bigl[12ka - (1-3\alpha )a^{-3\alpha}p_T\bigr]\,\,\,\,\, , \,\,\,\,\,  {\dot p}_{\Phi}=-12 N k \Phi \,\,\,\,\, , \,\,\,\,\,
{\dot p}_T=0
\end{array}
\label{classical}
\end{equation}

\noindent supplemented by the super-Hamiltonian constraint

\begin{equation}
\label{superconstraint}
- \frac{p_{a}^2}{24} + \frac{p_{\Phi}^2}{24} - 6ka^2 +6k{\Phi}^2 + a^{1-3\alpha }p_{T} = 0\hspace{0.3cm}.
\end{equation}
It follows that $\, p_T = \mbox{constant}$,  and in the conformal-time gauge ($ N=1$) we have

\begin{equation}
\label{classical2}
{\ddot \Phi} + k \Phi = 0 \,\,\,\,\, , \,\,\,\,\, {\dot T}= \,a^{1-3\alpha } \,\,\,\,\, , \,\,\,\,\,
{\ddot a}= - k a + \frac{1-3\alpha }{12} a^{-3\alpha}p_T
\hspace{0.3cm}.
\end{equation}

The cases $\, \alpha = -1\,$ and $\, \alpha = 0\,$ have been previously studied  in \cite{flavio2}.
 If $\, \alpha = 1/3$, the
 equation of motion  for the scale factor becomes identical to that satisfied by $\, \Phi\,$. If $\, \alpha\neq 1/3\,$,
equation (\ref{classical2}) for the scale factor corresponds to Newton's equation of motion  for  a particle of unit mass under
 a force whose potential is $\, V(a)= ka^2/2 + \ell a^{1-3\alpha} \,$, where
$\, \ell = - p_T / 12\,$ is a constant.
Making use of ``energy" conservation we find

\begin{equation}
t = \int \frac{da}{\sqrt{{\cal E} - ka^2 - \ell a^{1-3\alpha}}}
\hspace{0.2cm},
\end{equation}
where $\, {\cal E}\,$ is an integration constant. The above integral can be expressed in terms of elementary
functions in the cases of radiation $(\alpha=1/3)$, stiff matter ($\alpha =1$) and cosmic strings
($\alpha =-1/3$). In  other physically interesting cases such as dust $(\alpha=0)$, domain walls
$(\alpha=-2/3)$ and vacuum $(\alpha=-1)$ the integral is given by Jacobi elliptic functions.

\hspace{0.6cm}For future reference we solve the classical equations of motion in the radiation case
($\alpha=1/3$).
In the conformal-time gauge $(N=1)$  eqs.(\ref{classical}) are solved by
(the three lines below correspond to $k=-1,\, 0,\, 1$, respectively)
\begin{equation}
\begin{array}{lllll}
a(\tau) =& c_{1}\sinh\tau+c_{2}\cosh\tau\,\,\,\,\, ,
\,\,\,\,\, \Phi (\tau) = c_{3}\sinh\tau+c_{4}\cosh\tau \,\,\, ,\\
& \\
a(\tau) =& c_{1}\tau+c_{2}\,\,\,\,\, ,\,\,\,\,\, \Phi(\tau) = c_{3}\tau+c_{4}\,\,\, ,\\
& \\
a(\tau) =& c_{1}\sin\tau+c_{2}\cos\tau \,\,\,\,\, ,\,\,\,\,\,
\Phi (\tau) = c_{3}\sin\tau+c_{4}\cos\tau\,\,\, ,
\end{array}
\label{eq3}
\end{equation}

\noindent
where $c_{1}$, $c_{2}$, $c_{3}$,  $c_{4}$  and $p_{T}$ are integration constants. For $k=1$ both $a(\tau)$ and $\Phi (\tau)$ are
oscillating functions. The cases $k=-1$ and $k=0$ correspond to an ever-expanding Universe.

\section{Quantization}

\hspace{0.6cm}The  Wheeler-DeWitt quantization scheme  consists in promoting the canonical
momenta to operators according to
\begin{equation}
p_{a}\rightarrow -i\frac{\partial}{\partial a}\hspace{0.2cm},\hspace{0.2cm}p_{\Phi}\rightarrow -i\frac{\partial}{\partial \Phi}\hspace{0.2cm},\hspace{0.2cm}p_{T}\rightarrow -i\frac{\partial}{\partial T}\hspace{0.2cm},
\label{eq4}
\end{equation}

\noindent
and, with the corresponding super-Hamiltonian  operator $\hat {\cal{H}}$, forming the  Wheeler-DeWitt equation
\begin{equation}
{\hat{{\cal{H}}}}\Psi (a,\Phi,T)=0,
\label{eq5}
\end{equation}

\noindent
where $\Psi (a,\Phi,T)$ is called  the wave function of the Universe. In our case, the Wheeler-DeWitt
equation associated with eq.(\ref{superconstraint}) takes the  form of the Schr\"{o}dinger-like equation

\begin{equation}
\bigg(\frac{1}{24}\frac{{\partial}^2}{\partial a^2} - 6k a^2\bigg)\Psi(a,\Phi,T) - \bigg(\frac{1}{24}\frac{{\partial}^2}{\partial {\Phi}^2} -
 6k{\Phi}^2\bigg)\Psi(a,\Phi,T)= i \, a^{1-3\alpha}\hspace{0.1cm}\frac{\partial}{\partial T}
\Psi(a,\Phi,T).
\label{eq6}
\end{equation}

For the sake of convenience, let us perform the reparametrization
\begin{equation}
a=\frac{R}{\sqrt{12}}\hspace{0.2cm},\hspace{0.2cm}\Phi=\frac{\chi}{\sqrt{12}}\hspace{0.2cm},\hspace{0.2cm}\tau=- T\hspace{0.3cm},
\label{eq7}
\end{equation}
\noindent
which casts equation (\ref{eq6}) into the form
\begin{equation}
\left[\bigg(-\frac{1}{2}\frac{{\partial}^2}{\partial R^2} + \frac{k}{2}R^2\bigg)
-\bigg(-\frac{1}{2}\frac{{\partial}^2}{\partial {\chi}^2} + \frac{k}{2}{\chi}^2\bigg)\right]\Psi(R,\chi,\tau)
=i\,\bigg(\frac{R}{\sqrt{12}}\bigg)^{1-3\alpha} \,
\frac{\partial \Psi(R,\chi,\tau)}{\partial \tau}\, .
\label{eq8}
\end{equation}

\noindent
Eq.(\ref{eq8}) can be written as $i\partial
\Psi/\partial t ={\hat H} \Psi$ with a self-adjoint Hamiltonian operator  if the   inner
product is \cite{alvarenga}

\begin{equation}
(\Psi ,\Phi ) =  \int_{-\infty}^{\infty}d\chi\int_0^{\infty} dR\, R^{1-3\alpha} \,\Psi(R,\chi )^*\, \Phi (R, \chi )\,\,\, .
\label{eq9}
\end{equation}

\section{Static Quantum Wormholes}

\hspace{0.6cm}For the radiation case ($\alpha=1/3$) and $k=1$ the stationary solutions of
the  Wheeler-DeWitt equation  (\ref{eq8}) take the  form
$\Psi(R,\chi,\tau)=\varphi_n(R)\varphi_m(\chi)e^{-iE\tau}$ where $\varphi_n$ is the n-th eigenfunction of the  harmonic oscillator and $E=E_{nm}=n-m$, with
$n,m=0,1,2,\ldots$. The  wave functions
$\,\psi_{nm}(R,\chi) =
 \varphi_n(R)\varphi_m(\chi)\,$
form a  well-known  set of
wormhole quantum states \cite{Hawking,Garay1,Garay2,Mena}.

We now show that  static wormholes also exist for cosmic strings  and $k=1$. With $\alpha=-1/3$
the Wheeler-DeWitt equation (\ref{eq8}) becomes

\begin{equation}
\bigg(-\,\frac{1}{2}\frac{{\partial}^2}{\partial R^2} + \frac{1}{2}k R^2\bigg)\Psi(R,\chi,\tau) -\bigg(-\,\frac{1}{2}\frac{{\partial}^2}{
\partial {\chi}^2} + \frac{1}{2}k{\chi}^2\bigg)\Psi(R,\chi,\tau)=i\,\frac{R^2}{12} \hspace{0.1cm}
\frac{\partial}{\partial \tau} \Psi(R,\chi,\tau).
\label{eq10}
\end{equation}

\noindent Writing $\Psi = e^{-iE\tau}\, \psi$ we find

\begin{equation}
\frac{1}{2}\bigg[-\frac{\partial^2 \psi}{\partial R^2} + {k}R^2\psi\bigg] -
\frac{1}{2}\bigg[-\frac{\partial^2 \psi}{\partial {\chi}^2} + {k}{\chi}^2\psi\bigg] =
E \,\frac{R^2}{12}\psi\, .
\end{equation}

\noindent Applying  the method of separation of variables with $\psi (R,\chi ) = X(R)\, Y(\chi)$ we get

\begin{equation}
\frac{1}{2X}[-{X}'' + kR^2 X] - E\, \frac{R^2}{12} = \frac{1}{2Y}[-{Y}'' + k{\chi^2 Y}] = \frac{A}{2} \,\,\, ,
\end{equation}

\noindent where $A$ is a separation constant. Thus we have the following ordinary differential equations:

\begin{equation}
{X}'' + (A - kR^2
+ E\,\frac{R^2}{6}) X = 0 \,\,\, ,
\end{equation}

\begin{equation}
{Y}'' + (A - k\chi^2)Y = 0\,\,\,\, .
\end{equation}

\noindent With $k=1$ and  $A = 0$  these equations reduce to

\begin{equation}
\label{eqX}
{X}'' - \bigg( 1- \frac{E}{6}\bigg)R^2X  = 0 \,\,\, ,
\end{equation}

\begin{equation}
Y'' - \chi^2 Y = 0\,\,\, .
\end{equation}

\noindent The general solution to these equations which is regular everywhere and exponentially damped
for both
large scale factor and large scalar field is \cite{Hildebrand}

\begin{equation}
X(R)=R^{1/2}K_{1/4}\bigl(\sqrt{1-E/6}\, R^2/2\bigr)\,\,\, ,
\end{equation}

\begin{equation}
Y(\chi )=\vert {\chi}\vert ^{1/2} K_{1/4}(\chi^2/2)\,\,\, ,
\end{equation}

\noindent where  $E<6$ and  $K_{\nu}$ is the modified Bessel function of the second kind, of order $\nu$.

Thus, the static wormhole wave functions in this case are

\begin{equation}
\psi_E (R,\chi ) =N_E\,
(\vert\chi\vert R)^{1/2}
K_{1/4}(\chi^2/2)K_{1/4}\bigl(\sqrt{1-E/6}\, R^2/2\bigr)
\,\,\, ,
\label{wormcosmicstrings}
\end{equation}
\noindent with  $E<6$. These wave functions are normalizable in the inner product (\ref{eq9}).
The normalization
constant is found to be [Ref.\cite{Gradshteyn}, formulas 6.521(3) and 6.576(4)]

\begin{equation}
 N_E = \frac{2}{\pi^{3/2}}\, \sqrt{1-E/6}
\,\,\, .
\label{normalizationconstant}
\end{equation}
\noindent
Thus, in the case of a fluid of cosmic strings there exists a continuous family  of wormhole quantum
states. Because the Hilbert space of states is separable, the wave functions (\ref{wormcosmicstrings})
cannot be mutually ortogonal,
 just like the elements of the overcomplete set
of coherent states of the harmonic oscillator. Indeed,  we have

\begin{equation}
(\psi_{E^{\prime}}, \psi_E ) = \frac{4\,(1-E^{\prime}/6)^{3/8}\, (1-E/6)^{3/8}}
{[(1-E^{\prime}/6)^{1/4}\, + \,(1-E/6)^{1/4}]\, [(1-E^{\prime}/6)^{1/2}\, + \, (1-E/6)^{1/2}]}
\,\,\, .
\label{wormcosmicinnerproduct}
\end{equation}

\noindent The expectation value of the scale factor in  the state (\ref{wormcosmicstrings}) is

\begin{equation}
\langle R \rangle = \frac{2^{7/2}}{3\pi}\, \frac{\Gamma (1/4)^2}{(1-E/6)^{1/4}}
\,\,\, ,
\label{wormcosmicexpectation}
\end{equation}

\noindent which may be interpreted as the radius of the throat. Since $-\infty <E <6$,
static wormhole
universes with arbitrarily small or arbitrarily large throat radii are allowed. From equations (\ref{eq4}), (\ref{eq7})
and $\,\Psi = e^{-iE\tau }\, \psi\,$ it follows that $p_T=E$. For $E <6$ the classical equations of
motion (\ref{classical2}) with $\alpha = -1/3$
have no static solution for the scale factor. Thus, the quantum regime
in the state (\ref{wormcosmicstrings}) is such
that the classical evolution  of the universe is halted.
\section{Evolving Quantum Wormholes}

\hspace{0.6cm}In the case of cosmic strings, let us consider $E>6$. Setting $E= 6(1+\epsilon^2)$ with
$\epsilon>0$, a solution to eq.(\ref{eqX}) is

\begin{equation}
\label{eqXcosmic}
X(R)=R^{1/2}J_{1/4}\bigl(\epsilon\, R^2/2\bigr)\,\,\, ,
\end{equation}
where $J_{\nu}$ is the Bessel function of the first kind, of order $\nu$. The stationary wave functions

\begin{equation}
\psi_{\epsilon} (R,\chi ) =
(\vert\chi\vert R)^{1/2}
K_{1/4}(\chi^2/2)J_{1/4}\bigl(\epsilon\, R^2/2\bigr)
\label{stationarycosmic}
\end{equation}
do not represent physical states
because they  are not normalizable in the inner product (\ref{eq9}). Normalizable wave packets
may be constructed by writing

\begin{displaymath}
\Psi (R,\chi ,\tau)  =\int_0^{\infty}C(\epsilon ) e^{-iE\tau}\psi_{\epsilon} (R,\chi )d\epsilon
\end{displaymath}
\begin{equation}
\label{cosmicpacketintegral}
=e^{-6i\tau}
(\vert\chi\vert R)^{1/2}
K_{1/4}(\chi^2/2) \int_0^{\infty}C(\epsilon ) e^{-6i\tau\epsilon^2}
J_{1/4}\bigl(\epsilon\,R^2/2\bigr)d\epsilon
\,\,\, .
\end{equation}
The choice

\begin{equation}
\label{choicecosmic}
C(\epsilon )=\epsilon^{5/4}e^{-\gamma \epsilon^2}\,\,\, ,
\end{equation}
with $\gamma$ a positive constant, leads to [Ref.\cite{Gradshteyn}, formula 6.631(4)]

\begin{equation}
\Psi (R,\chi ,\tau)
\label{cosmicpacket}
=\frac{e^{-6i\tau}}{(\gamma +6i\tau )^{5/4}}\,
\vert\chi\vert^{1/2}
K_{1/4}(\chi^2/2) \, R\, \exp\left[ -\frac{R^4}{16(\gamma +6i\tau )}\right]
\,\,\, ,
\end{equation}
up to a normalization factor. The expectation value of the scale factor in the state (\ref{cosmicpacket}) is

\begin{equation}
\langle R \rangle (\tau )= \frac{2^{7/4}\sqrt{\pi}}{\Gamma (1/4)}\,\left(\frac{\gamma^2 + 36\tau^2}{\gamma}
\right)^{1/4}\,\,\, .
\label{cosmicpacketexpectation}
\end{equation}
Note that the mean radius of the Universe grows without bound in spite of the fact that the
spatial geometry of the Universe is
closed ($k=1$). Taking $\gamma$  sufficiently small, so that the wave packet is very well localized near $R=0$ at $\tau =0$,
the expectation value of the scale factor  becomes as large as one pleases immediately after $\tau =0$.
In other words, it is nearly certain that the Universe will have an enormous radius at any positive time.
This was called ``inflation without inflation" in \cite{Tipler,lemos3}.

Let us take up again the radiation case. With $\alpha = 1/3$ we can put equation (\ref{eq8}) in the form
\begin{equation}
i\hspace{0.15cm}\frac{\partial}{\partial \tau} \Psi(R,\chi,\tau)={\hat H}\Psi(R,\chi,\tau) \equiv ({\hat H}_R - {\hat H}_{\chi})\Psi(R,\chi,\tau)
\label{eq20}
\end{equation}

\noindent
where
\begin{equation}
{\hat H}_R = -\frac{1}{2}\frac{\partial^2}{\partial R^2} + \frac{k}{2}
R^2\,\,\,\,\, , \,\,\,\,\, {\hat H}_{\chi} =
-\frac{1}{2}\frac{\partial^2}{\partial \chi^2} + \frac{k}{2}
\chi^2\hspace{0.3cm}.
\label{eq110}
\end{equation}

\noindent The Hamiltonian operator  $\,{\hat H}\,$ is self-adjoint in the inner  product
\begin{equation}
(\Psi ,\Phi ) =  \int_{-\infty}^{\infty}d\chi\int_0^{\infty} dR\, \,\Psi(R,\chi )^*\, \Phi (R, \chi )
\label{eq11}
\end{equation}

\noindent
as long as its domain is restricted to those wave functions such that
$\Psi (0,\chi )=0$ or $\psi^{\prime}(0, \chi)=0$, where the prime means partial derivative with respect
to  $R$ \cite{lemos3}.
Except for this domain restriction, ${\hat H}$ is the difference between: Hamiltonians of two harmonic oscillators
($k=1$);  Hamiltonians of two free  particles ($k=0$); or  Hamiltonians of two inverted oscillators ($k=-1$).

\hspace{0.6cm} The propagator for  (\ref{eq20}) is
\begin{equation}
G(\chi , R ; \chi^{\prime}, R^{\prime};  \tau) = G_{R}(R; R^{\prime}; \tau)G^{*}(\chi; \chi^{\prime};  \tau),
\label{eq14}
\end{equation}
where $G$ is the usual harmonic-oscillator propagator.
As concerns   $G_{R}$,
some care must be exercised since $R$ takes values only on the
 half-line  $R\geq 0$. For the sake of simplicity, we consider only those wave functions that
satisfy the condition
$\Psi^{\prime} (0,\ \chi, \tau)=0$. For this boundary condition, the propagator is given by \cite{lemos3}
\begin{equation}
G_{R}(R,R^{\prime};\tau) =  G(R; R^{\prime}; \tau)+G(R; -R^{\prime}; \tau)\,\,\, ,
\label{eq15}
\end{equation}

\noindent where
\begin{equation}
G(x,x^{\prime};\tau) = \Biggl( \frac{\sqrt{k}}{2\pi i \sin\sqrt{k}\, \tau}\Biggr)^{1/2}
\exp\Biggl\{\frac{i\sqrt{k}}{2\sin\sqrt{k}\,\tau}\Bigl[ (x^2 + {x^{\prime}}^2)\cos\sqrt{k}\, \tau -
2x x^{\prime}\Bigr]\Biggr\}
\label{eq16}
\end{equation}

\noindent
is the  propagator for a one-dimensional   harmonic oscillator  with  mass $m=1$ and frequency $w=\sqrt{k}$.

The quantum dynamics of the models will be studied by following the time evolution of the initial  Gaussian
wave function

\begin{equation}
{\Psi}(\chi , R , 0)=\sqrt{\frac{4\sqrt{\sigma\beta}}{\pi}}\, e^{-\sigma R^2-\beta {\chi}^2}\,\,\, ,
\label{pacoteinicial}
\end{equation}
where $\, \sigma \,\mbox{and}\, \beta \,$ are
 positive  constants.
Making use of
\begin{equation}
\Psi (\chi , R , \tau) = \int_{-\infty}^{\infty}d\chi^{\prime} \int_{0}^{\infty}dR^{\prime}\,
G(\chi , R ; \chi^{\prime}, R^{\prime}; \tau) \, \Psi (\chi^{\prime} , R^{\prime} , 0)
\,\,\,
 ,
\label{eq21}
\end{equation}

\noindent
the wave  packet takes the form
\begin{displaymath}
\Psi (R,\chi,\tau) = \sqrt{\frac{4\sqrt{\sigma\beta}}{\pi}}\ \sqrt{\frac{k}{\cos^2(\sqrt{k}\tau)
\left[2\sigma \tan(\sqrt{k}\tau)-i\sqrt{k}\right]\left[2\beta \tan(\sqrt{k}\tau)+i\sqrt{k}\right]}}
\end{displaymath}
\begin{displaymath}
\times \exp\left\{
-\frac{i\sqrt{k}}{2\tan(\sqrt{k}\tau)}\left[1- \frac{i\sqrt{k}}{\cos^2(\sqrt{k}\tau)\left[2\beta \tan(\sqrt{k}\tau)+i\sqrt{k}
\right]}\right]{\chi}^2\right.
\end{displaymath}
\begin{equation}
\left. +\frac{i\sqrt{k}}{2\tan(\sqrt{k}\tau)}
\left[1+ \frac{i\sqrt{k}}{\cos^2(\sqrt{k}\tau)
\left[2\sigma \tan(\sqrt{k}\tau)-i\sqrt{k}\right]}\right]R^2\right\}\,\,\, .
\label{eq22}
\end{equation}

\section{The Bohm-de Broglie Interpretation}

\hspace{0.6cm}The Bohm-De Broglie interpretation
\cite{Holland} is an alternative interpretation of quantum mechanics  which allows the treatment of a unique system such as the Universe,
which cannot be repeatedly prepared in the same state as required by the standard statistical interpretation.
The  Bohm-De Broglie interpretation starts by writing the wave function of the Universe in the form
\begin{equation}
\Psi = \Theta \exp (i{\cal S}),
\label{eq25}
\end{equation}

\noindent
where  $\Theta$ and ${\cal S}$ are real functions. Then, inserting (\ref{eq25}) into the
Wheeler-DeWitt equation (\ref{eq6}) with $\alpha=1/3$, there results
\begin{equation}
\begin{array}{lll}
\frac{\displaystyle\partial {\cal S}}{\displaystyle\partial \tau} +&\frac{\displaystyle 1}{\displaystyle 2}
\left[\left(\frac{\displaystyle\partial {\cal S}}{\displaystyle\partial \chi}\right)^{2}-
\left(\frac{\displaystyle\partial {\cal S}}{\displaystyle\partial R}\right)^{2}\right]+V+Q=0 \, ,\\
\hspace{0.0cm} & \\
\frac{\displaystyle\partial \Theta}{\displaystyle\partial \tau} +&\left[\left(\frac{\displaystyle\partial
\Theta}{\displaystyle\partial \chi}\right)\left(\frac{\displaystyle\partial
{\cal S}}{\displaystyle\partial \chi}\right)-\left(\frac{\displaystyle\partial
\Theta}{\displaystyle\partial R}\right)\left(\frac{\displaystyle\partial {\cal S}}
{\displaystyle\partial R}\right)\right]+\frac{\displaystyle\Theta}
{\displaystyle 2}\left[\frac{\displaystyle\partial^2 {\cal S}}{\displaystyle\partial \chi^2}-
\frac{\displaystyle\partial^2 {\cal S}}{\displaystyle\partial R^2}\right]=0\, , \\
\end{array}
\label{eq26}
\end{equation}

\noindent
where $V$ and $Q$ are the classical and quantum potentials, respectively:
\begin{equation}
V = \frac{\displaystyle  1}{\displaystyle  2}k\left(\displaystyle R^2-\chi^2\right)\,\,\,\,\, , \,\,\,\,\,
Q = - \frac{\displaystyle  1}{\displaystyle 2\Theta}\left[\frac{\displaystyle\partial^2 \Theta}
{\displaystyle\partial R^2}-\frac{\displaystyle\partial^2 \Theta}{\displaystyle\partial \chi^2}\right]\, .
\label{eq27}
\end{equation}
Let us first examine  the case of vanishing spatial curvature ($k=0$).

\subsection{\bf The Flat Case}

The wave packet is obtained by taking the limit as $k\to 0$ of the wave function (\ref{eq22}), and it takes the form
\begin{equation}
\Psi(R,\chi,\tau) =\sqrt{4\frac{\sqrt{\sigma\beta}}{\pi}}\ {\Psi}_{1}(R,\tau)\ {\Psi}_{2}(\chi,\tau)\
\label{eq28}
\end{equation}

\noindent
where
\begin{equation}
\begin{array}{lll}
{\Psi}_{1}(R,\tau) =& \sqrt{\displaystyle\frac{1}{2\sigma \tau - i}}
\exp {\left\{\left(\displaystyle\frac{i}{2\tau}\, \frac{2\sigma \tau }{2\sigma
\tau - i}\right)R^2\right\}}\, ,\\
& \\
{\Psi}_{2}(\chi,\tau) =& \sqrt{\displaystyle\frac{1}{2\beta \tau + i}}
\exp{\left\{\left(\displaystyle-\frac{i}{2\tau}\, \frac{2\beta \tau }{2\beta \tau + i}
\right)\chi^2\right\}}.\\
\end{array}
\label{eq29}
\end{equation}

The wave function  (\ref{eq28}) may be  written in the form (\ref{eq25}) with the functions $\Theta$ and ${\cal S}$ given by
\begin{equation}
{\cal S} = \frac{2\sigma^2\tau}{1+4\sigma^2\tau^2}\,R^2 -
\frac{2\beta^2\tau}{1+4\beta^2\tau^2}\,\chi^2 + f_0(\tau)\, ,
\label{eq33}
\end{equation}
\begin{equation}
\Theta = g_0(\tau)
\exp{\left\{
-\displaystyle\frac{\sigma R^2}{1+4\sigma^2\tau^2} -
\displaystyle\frac{\beta\chi^2}{1+4\beta^2\tau^2}
\right\}}\, ,
\label{eq34}
\end{equation}
\noindent where $f_0$ and $g_0$ are  functions of $\tau$ alone that play
no role in the subsequent discussion.
From this we can find the Bohmian trajectories for the
scale factor  $a(\tau)$
and the scalar field $\Phi (\tau)$,
recalling that  $a$, $\Phi$, $R$ and  $\chi$ are related by  (\ref{eq7}), and also
that in the $\tau$-gauge one must take  $N=-1$ in Hamilton's equations (\ref{classical}).
The Bohmian trajectories are the solutions of the differential equations
\begin{equation}
p_{R} = \frac{\partial {\cal S}}{\partial R}\,\,\,\, , \,\,\,\,\,
p_{\chi} = \frac{\partial {\cal S}}{\partial \chi}\, .
\label{eq35}
\end{equation}

\noindent
Taking into account the previous remark on the lapse function, the classical equations of motion and the relations (\ref{eq7}), we get
\begin{equation}
\dot{R}(\tau) = \frac{4 \sigma^2\tau}{1+4 \sigma^2\tau^2}\, R(\tau).
\label{eq36}
\end{equation}

\noindent
An immediate integration yields
\begin{equation}
R(\tau) = A_{0}\sqrt{1+4\sigma^2\tau^2}\,\,\, ,
\label{eq37}
\end{equation}

\noindent
where $A_{0}$ is a positive constant of  integration.
There is no singularity since the Bohmian
trajectories $R(\tau)$ never reach $R=0$.

In a similar fashion, we find for the scalar field
\begin{equation}
\dot{\chi}(\tau) = \frac{4 \beta^2\tau}{1+4\beta^2\tau^2}\, \chi (\tau),
\label{eq38}
\end{equation}

\noindent
whose solution is
\begin{equation}
\chi(\tau) =B_{0}\sqrt{1+4\beta^2\tau^2}\,\,\, ,
\label{eq39}
\end{equation}

\noindent
where $B_0$ is a real constant of integration. The probability distribution of $\chi$
derived from (\ref{eq34}) is an even function of $\chi$. Thus, averaging over $B_0$ we obtain
the expectation value $\langle \chi \rangle =0$,
as it should.

The quantum potential $Q$ is found by inserting (\ref{eq34}) into (\ref{eq27}):
\begin{equation}
Q(R,\chi,\, \tau) = \frac{\sigma}{1+4\sigma^2\tau^2} - \frac{\beta}{1+4\beta^2\tau^2}
-\frac{2\sigma^2 R^2}{\left(1+4\sigma^2\tau^2\right)^2} +
\frac{2\beta^2\chi^2}{\left(1+4\beta^2\tau^2\right)^2}
\, .
\label{eq40}
\end{equation}

 The quantum force associated with the time evolution of the scale factor is

\begin{equation}
F^{(Q)}_{R} = -\frac{\partial Q}{\partial R} =
\frac{4\sigma^2R}{\left(1+4\sigma^2\tau^2\right)^2}\, .
\label{forcaquantica0}
\end{equation}

\noindent With the help of the trajectory (\ref{eq37}), the  quantum force $\, F_R\,$ can be expressed as a function of
$R$ only:
\begin{equation}
\label{quantumforce}
F^{(Q)}_R(R) = \frac{C}{R^3}\, ,
\end{equation}
\noindent where $C$ is a positive constant. This force is always
repulsive away from  $\, R=0\,$, is strongest when $R(\tau)$ is minimum  and is responsible for the  avoidance of the singularity in the quantum domain.

\subsection{\bf The Case of Positive Spatial Curvature}

\hspace{0.6cm}Now the wave function  (\ref{eq22}) takes the form
\begin{equation}
\Psi (R,\chi,\tau) = \sqrt{\frac{4\sqrt{\sigma\beta}}{\pi}}\ \Theta \exp\left(i{\cal S}\right)\, ,
\label{eq42}
\end{equation}

\noindent
with  ${\cal S}$ and $\Theta$ given by

\begin{equation}
\Theta = g_1(\tau)
\exp\left\{-\frac{\sigma R^2}{\cos^2\tau+4\sigma^2\sin^2\tau}
-\frac{\beta\chi^2}{\cos^2\tau+4\beta^2\sin^2\tau}\right\}
\label{eq42b}
\end{equation}
\noindent
and
\begin{equation}
{\cal S} =  f_1(\tau)+
\frac{R^2}{2\tan\tau}\left[1-\frac{1}{\cos^2\tau +4\sigma^2
\sin^2\tau}\right]-\frac{\chi^2}{2\tan\tau}
\left[1-\frac{1}{\cos^2\tau+4\beta^2\sin^2
\tau}\right]\, .
\label{eq42c}
\end{equation}
The Bohmian trajectories are determined by
\begin{equation}
\dot{R}(\tau) = \frac{1}{\tan\tau}\left[1-
\frac{1}{\cos^2\tau+4\sigma^2\sin^2\tau}\right]R(\tau),
\label{eq47}
\end{equation}
\begin{equation}
\dot{\chi}(\tau) =
\frac{1}{\tan\tau}\left[1-
\frac{1}{\cos^2\tau+4\beta^2\sin^2\tau}\right]\chi(\tau)
\label{eq48}.
\end{equation}
Integration yields
\begin{equation}
R(\tau) = A_+\sqrt{\cos^2 \tau +4\sigma^2\sin^2\tau }\,\,\, , \,\,\,
\chi(\tau) =  B_+\sqrt{\cos^2 \tau + 4\beta^2\sin^2 \tau}\,\, ,
\end{equation}
\noindent
where $A_+$ and $B_+$ are  constants of integration, with $A_+>0$.
Both $R$ and $\chi$ oscillate eternally between minimum and maximum values.

The quantum potential is
\begin{equation}
Q(R,\chi, \tau) = h(\tau )
-\frac{2\sigma^2R^2}{\left(\cos^2\tau+4\sigma^2\sin^2\tau\right)^2}+
\frac{2\beta^2\chi^2}{\left(\cos^2\tau+4\beta^2\sin^2\tau\right)^2}\, ,
\label{eq51}
\end{equation}
\noindent from which it is easily seen that the quantum force is again of the form (\ref{quantumforce}).

Only the interval $\,[0,\pi]\,$ is physically acceptable
for $\tau$ at the classical level, corresponding to the initial and final singularities. In the quantum
realm, however, $\tau$ may take values over the whole real line,  there being no singularity in virtue
of the repulsive character of the quantum force.
Differently from the flat case, in which only the quantum potential exists, in  the present case
($k=1$) the classical
an quantum potentials compete. However, as equation (\ref{quantumforce}) shows, the repulsive quantum force
is much stronger that the classical force near the singularity.

\subsection{\bf The Case of Negative Spatial  Curvature}

\hspace{0.6cm} The  results for $k=-1$ follow from those for $k=1$ by simply replacing
the trigonometric functions by the corresponding hyperbolic functions.
The Bohmian trajectories satisfy
\begin{equation}
\dot{R}(\tau) = \frac{1}{\tanh\tau}\left[1-
\frac{1}{\cosh^2\tau+4\sigma^2\sinh^2\tau}\right]R(\tau),
\label{eq56}
\end{equation}

\vspace{0.5cm}

\begin{equation}
\dot{\chi}(\tau) =
\frac{1}{\tanh\tau}\left[1-
\frac{1}{\cosh^2\tau+4\beta^2\sinh^2\tau}\right]
\chi(\tau),\label{eq57}
\end{equation}

\noindent
which are solved by

\begin{equation}
R(\tau) = A_-\sqrt{\cosh^2 \tau + 4\sigma^2\sinh^2 \tau}\,\,\, ,\,\,\,
\chi(\tau) = B_-\sqrt{\cosh^2 \tau + 4\beta^2\sinh^2 \tau}\, .
\end{equation}

\noindent with $A_->0$. Once again the quantum force as a function of $R$ is  given by (\ref{quantumforce}).

\section{Conclusion and Final Remarks}

\hspace{0.6cm}In this work we  quantized  Friedmann-Robertson-Walker cosmological models having for
matter content a perfect
fluid and a conformal scalar field. We found static wormhole wave functions in the cases of radiation
and cosmic strings. In the radiation case the  discrete set of wormhole quantum states was previously
known in the literature.
In the case of cosmic strings we found a new  continuous family
of static quantum  wormholes with arbitrary throat radii. The members of this  continuous set
of finite-norm wave functions
are not orthogonal to each other because the Hilbert state of states is separable. In this aspect
 they resemble
the overcomplete set of coherent states of the harmonic oscillator. For cosmic strings an evolving-wormhole
wave packet was constructed leading to a dynamically open universe, although the spatial geometry is closed.
If the initial state of the Universe is sufficiently localized near the classical singularity, the scale
factor is enormously large immediately after the initial instant, a phenomenon described in the literature as
``inflation without inflation".
In the radiation case,
with the help of the exact propagator
we  constructed  wave packets that
are examples of dynamic quantum wormholes. The behaviour of the scale factor was studied
according to the Bohm-de Broglie
interpretation of quantum mechanics.
In all three cases ($k=0,\pm 1$)
the singularity is prevented by a quantum force inversely proportional to the cube of the
scale factor. Also, in all three cases, the integral $\int_{-\infty}^t d\tau^{\prime}/R(\tau^{\prime})$
is divergent,  so that for the states considered the models have no particle horizon.

\section*{Acknowledgment}

The authors thank the Conselho Nacional de Desenvolvimento Cient\'{\i}fico e Tecnol\'ogico (CNPq), Brazil,
for finantial support. Special thanks are due to F. G. Alvarenga and J. C. Fabris for discussions and for
reading a preliminary version of the manuscript.


\begin{thebibliography}{60}
\bibitem{DeWitt} B. S. DeWitt, Phys. Rev. {\bf 160}, 1113 (1967).
\bibitem{Hawking} S. W. Hawking and D. B. Page, Phys. Rev. {\bf D42}, 2655 (1990).
\bibitem{flavio2} N. A. Lemos and F. G. Alvarenga, Gen. Relat. Grav. {\bf 31}, 1743 (1999).
\bibitem{alvarenga} F. G. Alvarenga, J. C. Fabris, N. A. Lemos  and G. A. Monerat,
Gen. Relat. Grav.  {\bf 34}, 651 (2002).
\bibitem{Schutz} B. F. Schutz,  Phys. Rev. {\bf D2}, 2762 (1970); {\bf D4}, 3559 (1971).
\bibitem{Rubakov} V. G. Lapchinskii and V. A. Rubakov, Theor. Math. Phys. {\bf 33}, 1076 (1977).
\bibitem{lemos4} N. A. Lemos, Class. Quantum Grav. {\bf 8}, 1303 (1991).
\bibitem{Peleg} J. Feinberg and Y. Peleg, Phys. Rev. {\bf D52}, 1988 (1995).
\bibitem{Garay1} L. J. Garay, Phys. Rev. {\bf D44}, 1059 (1991).
\bibitem{Garay2} L. J. Garay, Phys. Rev. {\bf D48}, 1710 (1993).
\bibitem{Mena} G. A. Mena Marug\'an, Phys. Rev. {\bf D50}, 3923 (1994).
\bibitem{Hildebrand} F. B. Hildebrand, {\it Advanced Calculus for Applications} (Prentice-Hall, Englewood
Cliffs, NJ, 1976), Section 4.10.
\bibitem{Gradshteyn} I. S. Gradshteyn and I. M. Ryzhik, {\it Tables of Integrals, Series and Products} (Corrected
and Enlarged Edition, Academic, New York, 1980).
\bibitem{Tipler} F. J. Tipler, Phys. Rep. {\bf 137}, 231 (1986).
\bibitem{lemos3} N. A. Lemos, J. Math. Phys. {\bf 37}, 1449 (1996).
\bibitem{Holland} P. R. Holland, {\it The Quantum Theory of Motion} (Cambridge University
Press, Cambridge, 1993).
\end{thebibliography}
\end{document}